\documentstyle[prl,floats,twocolumn,aps,epsf,psfrag,amssymb]{revtex}

\tighten
\begin{document}
\draft
\twocolumn[\hsize\textwidth\columnwidth\hsize\csname @twocolumnfalse\endcsname
\title{Universal teleportation with a twist}

\author{Samuel L.\ Braunstein,${}^{1,2}$
Giacomo M.\ D'Ariano,${}^{3}$ G.\ J.\ Milburn${}^{4}$, and
Massimiliano F. Sacchi${}^{3}$}
\address{${}^1$SEECS, University of Wales, Bangor LL57 1UT, UK}
\address{${}^2$Hewlett-Packard Labs, Mail Box M48, Bristol BS34 8QZ, UK}
\address{${}^3$INFM, Unit\`a di Pavia, Dipartimento di Fisica
'A. Volta', Universit\`a di Pavia, via Bassi 6 -- I-27100 Pavia, ITALY}
\address{${}^4$Department of Physics, The University of Queensland,
QLD 4072 Australia}

\date{\today}
\maketitle

\begin{abstract}
We give a transfer theorem for teleportation based on {\it twisting\/}
the entanglement measurement. This allows one to  say what
local unitary operation must be performed to complete the teleportation
in any situation, generalizing the scheme to include overcomplete
measurements, non-abelian groups of local unitary operations
(e.g., angular momentum teleportation), and the effect of
non-maximally entangled resources.
\end{abstract}
\vspace{3ex}
]

One of the most profound results of quantum information theory is
the discovery of quantum teleportation protocols
\cite{Bennett93,vaidman94,Braunstein98,GC}.
Teleportation is the disembodied transport of quantum states between
subsystems through
a classical communication channel requiring a shared resource of
entanglement. The demonstration of teleportation elevates entanglement from
a perennial theoretical chestnut to a practical
resource. The details of protocols for teleportation may vary;
specification of subsystems, the shared entangled state, and
the description of joint measurements at the sender (Alice) or receiver (Bob).
For example already there have been several experimental implementations
of teleportation \cite{Zeilinger97,Demartini98,Furasawa98} and  other
protocols have been proposed \cite{Milburn99}.
We show in this paper that all teleportation schemes can be cast in a
common form with generalized (overcomplete) measurements
and which enables us to identify the local unitary operations required to
complete a teleportation scheme.

Let us start by recalling a maximally entangled
state in ${\cal H}\otimes{\cal H}$
\begin{equation}
|\Psi\rangle\!\rangle = \frac{1}{\sqrt{d}}\sum_n e^{-i\phi_n}
|n\rangle\otimes |n\rangle \;,
\end{equation}
where $\{ |n\rangle \}$ is any basis in ${\cal H}$ and $d$ is the
dimension of ${\cal H}$ (the infinite dimensional case will be
considered later). In the following we will adopt the
notation: use double ket (bra) $|\cdots\,\rangle\!\rangle$
to denote vectors in ${\cal H}\otimes{\cal H}$ and customary single
ket (bra) for vectors in ${\cal H}$.

It is now well known that we may span the set
of {\it all\/} maximally entangled states by local unitary
operations.
Therefore, it is sufficient to consider
local unitary operators acting only on one space.
In what follows we shall use a {\it twist\/} operation which swaps
a pair of particles. Here we introduce a `democratic' notation so that
given a pair of systems in a state
\begin{equation}
|\!\stackrel{\rightarrow}{\Psi}\rangle\!\rangle =
\sum_{jk}c_{jk}|j\rangle\otimes|k\rangle \;,
\end{equation}
the twisted/swapped version is denoted by
\begin{equation}
|\!\stackrel{\leftarrow}{\Psi}\rangle\!\rangle =
\sum_{jk}c_{jk}|k\rangle\otimes|j\rangle \;,
\end{equation}
and {\it vice-versa} 
$|\!\!\stackrel{\leftarrow}{\Psi}\rangle\!\rangle\leftrightarrow
|\!\!\stackrel{\rightarrow}{\Psi}\rangle\!\rangle$ . The
generalization to mixed states follows trivially. A final piece of notation
we introduce is the {\it transfer\/} operator
\begin{equation}
{\cal T}_{ba} = \sum_n |n\rangle_b \,{}_a\langle n| \;.
\end{equation}

We now give several identities which succinctly describe teleportation.
For an arbitrary {\it maximally\/} entangled state
$|\!\stackrel{\leftarrow}{\Psi}\rangle\!\rangle$ one can easily show that
\begin{equation}
\langle\!\langle \stackrel{\rightarrow}{\Psi}\!| \otimes \openone\;\;
\openone\otimes|\!\stackrel{\leftarrow}{\Psi}\rangle\!\rangle
= {1\over d}{\cal T}_{31} \label{transfer} \;.
\end{equation}
Because this notation is rather compact, we shall rewrite this equation\
labelling the particle numbers of each of the states and operators
involved, which gives
\begin{equation}
{}_{12}\langle\!\langle \stackrel{\rightarrow}{\Psi}\!| \otimes
{\openone}_3 \;\; {\openone}_1 \otimes
|\!\stackrel{\leftarrow}{\Psi}\rangle\!\rangle_{23}
= {1\over d}{\cal T}_{31} \;.
\end{equation}
We feel that these equations are more pleasing and equally unambiguous
without particle labels \cite{notation}.

Variations on the identity (\ref{transfer}) are
\begin{equation}
\langle\!\langle \stackrel{\rightarrow}{\Psi}\!| \otimes \openone\;\;\;
|\phi\rangle \otimes |\!\stackrel{\leftarrow}{\Psi}\rangle\!\rangle
= {1\over d}|\phi\rangle \label{entId} \;,
\end{equation}
where $|\phi\rangle$ is an arbitrary (unknown) quantum state; also
\begin{equation}
\openone \otimes \langle\!\langle \stackrel{\rightarrow}{\Psi}\!|
\otimes \openone\;\;\;
|\Phi\rangle\!\rangle \otimes
|\!\stackrel{\leftarrow}{\Psi}\rangle\!\rangle
= {1\over d}|\Phi\rangle\!\rangle \;,\label{swap}
\end{equation}
and similarly
\begin{equation}
\openone \otimes \langle\!\langle \stackrel{\rightarrow}{\Psi}\!|
\otimes \openone\;\;\;
|\!\stackrel{\leftarrow}{\Psi}\rangle\!\rangle \otimes
|\Phi\rangle\!\rangle
= {1\over d}|\Phi\rangle\!\rangle \;,\label{swap2}
\end{equation}
which will correspond to {\em entanglement swapping} for an arbitrary unknown
(entangled) two-mode state $|\Phi\rangle\!\rangle$. Some other trivial
variations of these identities are
\begin{equation}
\langle\!\langle \stackrel{\leftarrow}{\Psi}\!| \otimes \openone\;\;
\openone \otimes |\!\stackrel{\rightarrow}{\Psi}\rangle\!\rangle =
{1\over d}{\cal T}_{31} \;.\label{T31}
\end{equation}
and
\begin{equation}
\openone\otimes\langle\!\langle \stackrel{\rightarrow}{\Psi}\!| \;\;
|\!\stackrel{\leftarrow}{\Psi}\rangle\!\rangle\otimes\openone =
\openone\otimes\langle\!\langle \stackrel{\leftarrow}{\Psi}\!| \;\;
|\!\stackrel{\rightarrow}{\Psi}\rangle\!\rangle\otimes\openone
= {1\over d}{\cal T}_{13} \;,
\end{equation}
and other identities analogous to Eqs.~(\ref{entId}) and~(\ref{swap}) follow.

Let us see how these identities allow us to understand teleportation.
Start with an unknown state and a shared arbitrary maximally entangled
resource $|\phi\rangle \otimes |\Phi\rangle\!\rangle$. Perform a measurement
on the first two subsystems yielding a maximally entangled result
$|\!\!\stackrel{\rightarrow}{\Psi}\rangle\!\rangle$. We emphasize that this
measurement may be complete or overcomplete. Information about which
entangled state was found by Alice is transmitted to Bob. To complete the
teleportation protocol Bob must convert $|\Phi\rangle\!\rangle$ into the
twisted version of the entangled state actually found by Alice, i.e.,
$|\!\stackrel{\leftarrow}{\Psi}\rangle\!\rangle$. This conversion involves a
local unitary operation which now leaves us with the situation described
by Eq.~(\ref{entId}). Using it shows that the initial unknown state at
Alice's end has been successfully transferred to Bob.

At this point it is worthwhile stepping back and looking at what this
teaches us about quantum teleportation. In the ideal case Alice and Bob
must share a maximally entangled state and Alice must be able to perform
a measurement which yields a maximally entangled state. The details of the
measurement, for example, whether it involves projection measurements or
a POVM is unimportant. The reconstruction operation only relies on Bob
being able to locally convert his shared entanglement into the swapped
version that Alice found. But this is a general feature of maximally
entangled states. In fact, it lies at the heart of several other quantum
communication protocols. In quantum dense coding this ability allows
us to encode the square as many orthogonal states as are supported
by the Hilbert space we are acting on \cite{Bennett92}. This yields
potentially a
doubled channel capacity. Similarly, in any scheme which tries to
implement bit commitment, the freedom to locally convert any maximally
entangled state to any other allows Alice to cheat with impunity
\cite{Mayers97,Lo97}.
Now we have shown that this same freedom also drives the quantum
teleportation protocol. This commonality improves our understanding of
the ways in which the manipulation of shared entanglement may be used.

In order to interpret the vector
$|\!\!\stackrel{\rightarrow}{\Psi}\rangle\!\rangle$
as the result of a measurement, we need an (over)complete set of maximally
entangled vectors. This can be easily achieved by having the unitary
operator $U\equiv U(g)$ as an element of a group ${\mathbf G}=\{ g\}$ of
transformations $g$ with unitary irreducible representation (UIR) $U(g)$ on
Alice's Hilbert space ${\cal H}$. Then, for {\it any\/} maximally entangled
state $|\Psi\rangle\!\rangle$, one has the identity
\begin{equation}
\int_{\mathbf G} \mbox{d} g\;\; U(g)\otimes\openone\;\;
|\Psi\rangle\!\rangle\langle\!\langle \Psi|\;\;
U^\dagger(g)\otimes \openone ={1\over d}\openone\otimes \openone
\;,\label{whow}
\end{equation}
which easily follows from the identity (Schur's lemma)
\begin{eqnarray}
\int_{\mathbf G} \mbox{d} g\;\; U(g)\, A\, U^\dagger(g)
= \mbox{Tr}(A)\, \openone\;,
\end{eqnarray}
which holds for any operator $A$ on ${\cal H}$. The invariant
measure $\mbox{d} g$ is normalized as
\begin{eqnarray}
\int_{\mathbf G}\mbox{d}g\;\;|\langle
u|U(g)|v\rangle|^2=1\;,\label{norm}
\end{eqnarray}
which is true for any pair of normalized vectors $|u\rangle$, and
$|v\rangle$ due to the irreducibility of the representation, (assuming
square integrable UIR for simplicity). Eq.~(\ref{whow})
means that the set of vectors
\begin{eqnarray}
|\!\stackrel{\rightarrow}{\Psi}_g
\rangle\!\rangle\equiv U(g)\otimes\openone\,
|\Psi \rangle\!\rangle\;,\qquad g\in{\mathbf G}\label{pom}
\end{eqnarray}
make a (generally not orthogonal) POVM that represents a measurement on
${\cal H}\otimes{\cal H}$ with result $g$. The measurement correlates
Alice's Hilbert space with the entangled resource. Alice gets the
result $g$ and communicates it to Bob classically, and as already
mentioned Bob converts his shared entanglement $|\Phi\rangle\!\rangle$
into the twisted version of the entangled state found by Alice, i.e.,
$|\!\stackrel{\leftarrow}{\Psi}_g\rangle\!\rangle=\openone\otimes U(g)
|{\Psi}\rangle\!\rangle$.
For each result $g$ the state $|\phi\rangle$ is teleported according
to the overall transformation
\begin{equation}
\langle\!\langle \stackrel{\rightarrow}{\Psi}_g\!| \otimes \openone\;\;\;
|\phi\rangle \otimes |\!\stackrel{\leftarrow}{\Psi}_g\rangle\!\rangle
={1\over d}|\phi\rangle \;,\label{entIdg}
\end{equation}

For discrete groups the sum replaces the integral over ${\mathbf
G}$. Mathematically, Eq.~(\ref{entIdg}) represents a {\em pure
instrument} \cite{dy}, which describes the state reduction depending on the
outcome $g$ of the measurement, and sends a pure state into a pure
state. In the general case such an instrument has the form
\begin{eqnarray} \frac{\Omega _x
|\phi \rangle}{|\! |\Omega _x |\phi \rangle |\! |} =|\phi _x \rangle\;,
\end{eqnarray}
where $x$ is the measurement outcome and
$|\phi _x \rangle $ is the state conditioned by the result $x$. The
case of teleportation is peculiar because the conditioned state is
identical to the original one, independent of the measurement
outcome, and on the other hand it is ``teleported'' to another
space. In such a scenario the teleportation map should be regarded in
the following way
\begin{eqnarray} \frac{\Omega_g
|\phi \rangle}{|\! |\Omega _g |\phi \rangle |\! |} =
{\cal T}_{31}|\phi \rangle\;,
\end{eqnarray}
where $\Omega_g=\langle\!\langle \stackrel{\rightarrow}{\Psi}_g\!|
\otimes \openone\;\;\openone\otimes
|\!\stackrel{\leftarrow}{\Psi}_g\rangle\!\rangle
\equiv\frac 1d{\cal T}_{31}$.
Notice that Eq.~(\ref{whow}) has the relevant feature that phase factors
in the group composition law can be neglected. In mathematical terms
this means that if the unitary representation is of the ``projective''
form
\begin{eqnarray}
U(g)U(g')=c(g,g')\,U(gg')\;,
\end{eqnarray}
where $c(g,g')$ is a phase factor---a so called {\em cocycle}
\cite{cocy}---then, because of the peculiar form of Eq.~(\ref{whow})
the phase factor $c(g,g')$ can be dropped.

The original case of Ref.~\onlinecite{Bennett93} corresponds to the group of
the four Pauli matrices $\{\openone,
\sigma_x\,,\sigma_y\,,\sigma_z\}$, which is a projective
representation of the abelian dihedral group $D_2$ of $\pi$-rotations
around three perpendicular axes. Notice that even though the
projective representation is non-abelian
(i.e. $\sigma_x\sigma_y=i\sigma_z =-\sigma_y\sigma_x $) the
represented group is abelian ($R_xR_y=R_yR_x=R_z$, $R_\alpha $ denoting a
$\pi$-rotation around the $\alpha=x,y,z$ axis).  

The generalization to dimension $N$ in Ref.~\onlinecite{Bennett93} is again a
projective representation of an abelian group, namely
${\mathbb Z}_N \times {\mathbb Z}_N$, which is the group of
discrete translations on a lattice embedded in a torus. The
representation of the group given in Ref.~\onlinecite{Bennett93} is
\begin{eqnarray}
U(n,m)=\sum_k e^{2\pi ikn/N}|k \rangle \langle k\oplus m| \;,
\end{eqnarray}
which satisfies the composition law
\begin{eqnarray}
U(n,m)\, U(n',m')=e^{2\pi i mn'/N}U(n\oplus n',m\oplus m')\;,
\end{eqnarray}
where $n\oplus n'$ denotes summation mod $N$.

\par For infinite dimensional Hilbert spaces the POVM related to
Eq.~(\ref{pom}) generally needs to be expressed in terms of
unnormalizable vectors. We rewrite Eq.~(\ref{pom}) as follows
\begin{eqnarray}
|\!\stackrel{\rightarrow}{\Theta}_g
\rangle\!\rangle\equiv U(g)\otimes\openone\,
\sum_n |n \rangle \otimes |n \rangle ,\qquad g\in{\mathbf G}\label{pom2}
\label{schmidt}
\end{eqnarray}
(In general, the variable $n$ may be continuous. In this case the
sum would be replaced by an integral.) Moreover, one needs to consider
non-maximally entangled states
\begin{eqnarray}
|\Psi(\lambda)\,\rangle\!\rangle = \sum_n c_n(\lambda)\,
|n\rangle \otimes | n\rangle \;,
\end{eqnarray}
which depend on a physical parameter $\lambda\in[0,1)$ (e.g., this could
be a down-conversion gain) such that the state becomes maximally entangled
in the limit of $\lambda\to 1$ with
$\lim_{\lambda\to 1} |c_{n+1}(\lambda)/c_n(\lambda)|=1$. Then we introduce
the {\it distortion\/} operator
\begin{eqnarray}
{\cal D}(\lambda)=\sum_n c_n(\lambda)\,|n\rangle\langle n|\;,
\end{eqnarray}
and Eq.~(\ref{entIdg}) now becomes
\begin{eqnarray}
\langle\!\langle \stackrel{\rightarrow}{\Theta }_g | \otimes \openone\;\;
\openone \otimes |\!\stackrel{\leftarrow}{\Psi}_{g'}
\!(\lambda)\,\rangle\!\rangle =
U(g')\,{\cal D}(\lambda)\,U^{\dag}(g)\,{\cal T}_{31} \;,\label{T31d}
\end{eqnarray}
where
\begin{equation}
|\!\stackrel{\leftarrow}{\Psi}_g\!(\lambda)\rangle\!\rangle =
 \openone \otimes U(g)\; |\Psi(\lambda)\,\rangle\!\rangle\;.
\end{equation}

The teleportation map is achieved for $g'=g$ in the limit of 
$\lambda \rightarrow 1$ as follows
\begin{eqnarray}
\lim _{\lambda \rightarrow 1}
\frac{U(g)\,{\cal D}(\lambda)\,U^{\dag}(g)\,{\cal
T}_{31}|\phi \rangle }{|\! |U(g)\,{\cal D}(\lambda)\,U^{\dag}(g)\,{\cal
T}_{31}|\phi \rangle|\! |}=|\phi \rangle \;.
\end{eqnarray}
The continuous variables teleportation of Ref.~\onlinecite{Braunstein98} is an
example of infinite dimensional teleportation. The group is the
Weyl-Heisenberg group of displacement operators $D(z)=e^{za^\dag -\bar
z a}$ (where $[a,a^{\dag}]=1$ for the harmonic oscillator algebra) with
composition
law $D(z)D(w)=e^{i\mbox{{\footnotesize Im}}(z\bar w)}D(z+w)$. Notice that
this is just
a projective representation of the abelian group of translations on
the complex plane. Eq.~(\ref{norm}) reads
$\int _{\mathbb C}\frac {d^2 z}{\pi }\, e^{-|z|^2}=1$
by taking $|u \rangle =|v \rangle =|0 \rangle$ ($|0 \rangle $
denoting the vacuum for $a$). The entangled state is just the
downconversion of the vacuum
\begin{eqnarray}
|\Psi (\lambda )\,\rangle\!\rangle =\sqrt{1-\lambda ^2}\sum_{n=0}^\infty
\lambda ^n |n \rangle \otimes |n \rangle \;,
\end{eqnarray}
(a phase factor for $\lambda$ can always be included into the basis
definition). For $\lambda < 1$ one has the teleportation map with
distortion 
\begin{eqnarray}
\frac{\Omega _z^{(\lambda )}|\phi \rangle }
{|\! |\Omega _z^{(\lambda )}|\phi \rangle|\! |}=|\phi _z^{(\lambda )}\rangle\;,
\end{eqnarray}
where $\Omega_z^{(\lambda )}=\langle\!\langle
\stackrel{\rightarrow}{\Theta}_z |
\otimes \openone\;\;\openone\otimes
|\!\stackrel{\leftarrow}{\Psi}_z\! (\lambda)\,\rangle\!\rangle$, with
$|\!\stackrel{\leftarrow}{\Psi}_z\! (\lambda)\,
\rangle\!\rangle=\openone \otimes D(z)
|\Psi (\lambda)\,\rangle\!\rangle$, and
$|\!\stackrel{\rightarrow}{\Theta}_z\,
\rangle\!\rangle=D(z)\otimes\openone
|\Theta\,\rangle\!\rangle$, the latter being the orthogonal POVM
corresponding to the eigenvectors of the heterodyne photocurrent \cite{hete}.

Teleportation for infinite dimensional Hilbert spaces is not restricted to
maximally entangled states based on
decomposition in Eq. (\ref{schmidt}). We can define teleportation { \it
filters} that only teleport part of the Hilbert
space \cite{Lo99}. An example is the entangled state that results from two
harmonic oscillator coherent states
$|\alpha\rangle\otimes|\beta\rangle$,  through the unitary transformation
$U_K=\exp(-i\pi a^\dagger ab^\dagger b)$ where $a,b$ are the annihilation
operators. The resulting state is
\begin{eqnarray}
|\Pi\rangle & = &
|\alpha\rangle\otimes|\beta_+\rangle+|-\alpha\rangle\otimes|\beta_-\rangle\\
  & = &
|\alpha_+\rangle\otimes|\beta\rangle+|\alpha_-\rangle\otimes|-\beta\rangle
\end{eqnarray}
where $|z_\pm\rangle=|z\rangle\pm|-z\rangle$ , which are sometimes called
cat states and are parity eigenstates (we have
ignored normalization). With this entangled resource/measurement we can
only teleport states that lie in the relevant two
dimensional parity subspace of the entire Hilbert space.

\par The universal scheme in the present letter allows teleportation
through entangled measurements based on non-abelian groups, which has
never been considered yet. The simplest case is angular momentum
teleportation. We parameterize the group representation matrices as
$U(g)=\exp(i\varphi\vec J\cdot\vec n)$, where $\varphi\in[0,2\pi)$
\cite{note}, $\vec n$ is a unit vector $|\vec n|^2=1$ on a sphere, and
$J_{\alpha}$ are customary angular momentum operators. With such a
parameterization the invariant measure is
$\mbox{d} g= \mbox{d}\vec n\,\sin^2(\varphi/2)\,\mbox{d}\varphi/8\pi$.
The teleportation map is then
\begin{equation}
\langle\!\langle \stackrel{\rightarrow}{\Psi}_{\varphi,\vec n}\!|
\otimes \openone\;\;\;
|\phi\rangle \otimes |\!\stackrel{\leftarrow}{\Psi}_{\varphi,\vec n}
\rangle\!\rangle
={1\over{2J+1}}|\phi\rangle \;,\label{entIdrot}
\end{equation}
where
\begin{eqnarray}
|\!\stackrel{\leftarrow}{\Psi}_{\varphi,\vec n}
\rangle\!\rangle\equiv \openone\otimes e^{i\varphi\vec J\cdot\vec n}\,
|\Psi\rangle\!\rangle\;,
\end{eqnarray}
for a fixed maximally entangled state $|\Psi\rangle\!\rangle$.

In this paper we have presented the essential mathematical description of
how entanglement plus local measurement and unitary
transformation enables teleportation. Dense coding \cite{Bennett92} can be
given a similar description, however the role played
by the classical and quantum information channels is interchanged (see
figure 1). Both schemes rely on the ability to map
shared entanglement to shared entanglement through local unitary
transformations. We are thus able to see the common role of
local entanglement manipulation in quantum communication protocols.

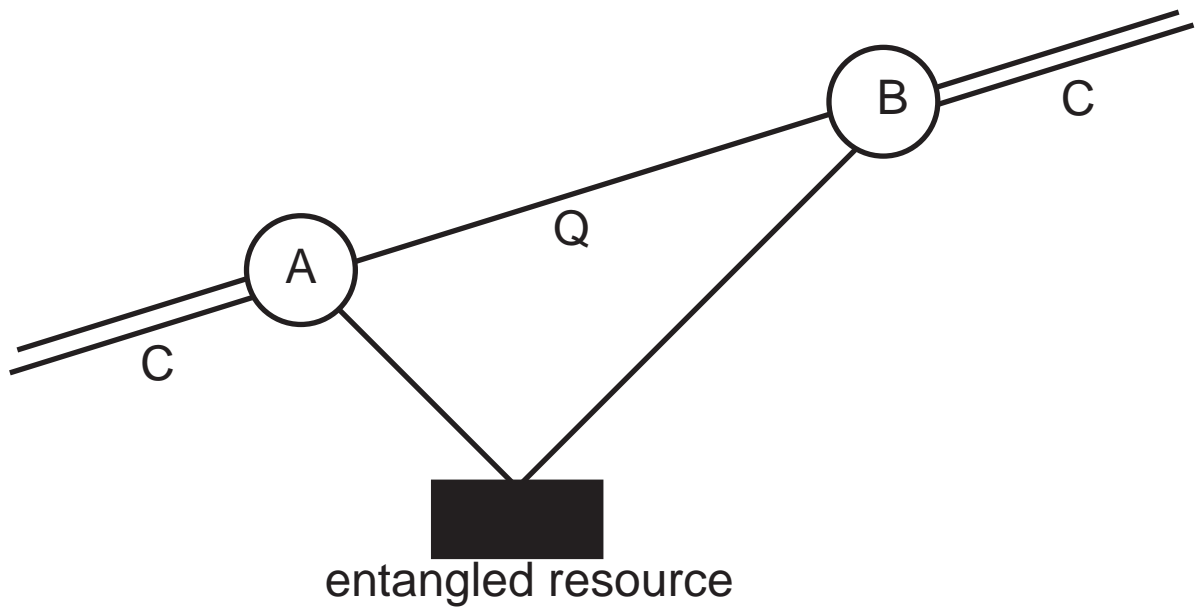
\begin{figure}
\caption{A schematic representation depicting how both teleportation and
dense coding use an entangled resource and classical communication. Time
runs vertically and space horizontally. A single line represents a quantum
state sent over a quantum (noiseless) channel (Q), a double line
represents classical information sent over an ordinary classical
communication channel  (C). In dense coding the quantum and classical
channels are interchanged from that for teleportation. It is already well
known that the steps in each protocol converting quantum to classical
information (mediated by shared entanglement) involve common Bell state
measurements. In this paper, we have furthermore shown that those steps
converting classical to quantum information (mediated by shared
entanglement) also operate on a common principle: one maximally entangled
state may be converted to any other by one-sided (local) operations.}
\label{fig1}
\end{figure}

\begin{references}
\bibitem{notation}The identity operators appearing in this expression
should be thought of as tensor slots with the first slot for bras 
(linear functionals) in ${\cal H}_3$ and the second slot for kets (vectors) 
in ${\cal H}_1$. When both slots are filled the operator defines a linear 
functional on ${\cal H}_1\otimes{\cal H}_3$ given by
$$
\mbox{}_3\langle f|{\cal T}_{31}|g\rangle_1=\mbox{}_3\langle
f|g\rangle_3=\mbox{}_1\langle f|g\rangle_1.
$$
\bibitem{cocy}The cocycle must satisfy the Jacobi associativity
constraints, namely that $c(gh,l)\, c(g,h)=c(g,hl)\, c(h,l)$ and
$c(g,g^{-1})=c(g,e)=1$, for $g,h,l\in{\mathbf G}$, $e$ being the identity
element.
\bibitem{note}Due to the particular form of Eq.~(\ref{whow}), the $-1$
factor of a $2\pi$ rotation for half-odd spins is canceled between the
bra and ket. Therefore, it is sufficient to consider $\psi\in[0,2\pi)$
instead of  $\psi\in[0,4\pi)$.
\end{references}
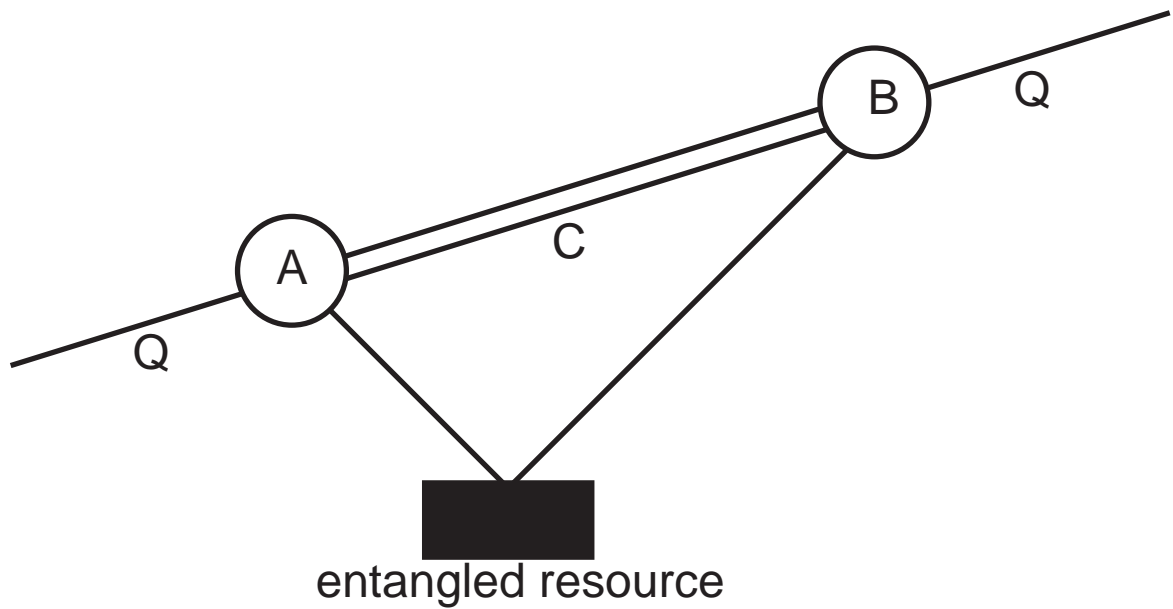
\end{document}